\documentclass[sn-mathphys,Numbered,iicol]{sn-jnl}

\usepackage{graphicx}%
\usepackage{multirow}%
\usepackage{amsmath,amssymb,amsfonts}%
\usepackage{amsthm}%
\usepackage{mathrsfs}%
\usepackage[title]{appendix}%
\usepackage{xcolor}%
\usepackage{textcomp}%
\usepackage{manyfoot}%
\usepackage{booktabs}%
\usepackage{algorithm}%
\usepackage{algorithmicx}%
\usepackage{algpseudocode}%
\usepackage{listings}%

\usepackage{latexsym}
\usepackage{dcolumn}
\usepackage[utf8]{inputenc}
\usepackage[T1]{fontenc}
\usepackage{tabularx}
\usepackage{hyperref}
\usepackage{orcidlink}
\usepackage{color}
\usepackage{placeins}
\usepackage{float}
\usepackage{gensymb}
\usepackage[separate-uncertainty=true]{siunitx}
\usepackage{soul}

\usepackage{cleveref}
\usepackage{subfigure}
\usepackage{subcaption}

\usepackage{array}
\newcolumntype{?}{!{\vrule width 1.5pt}}

\makeatletter
\def\hlinewd#1{%
\noalign{\ifnum0=`}\fi\hrule \@height #1 %
\futurelet\reserved@a\@xhline}
\makeatother

\DeclareSIUnit\angstrom{\text{Å}}

\renewcommand{\citet}[1]{Ref.~\cite{#1}}

\begin{document}

\title{Influence of bending parameters on crystalline undulator radiation peak stability for 530~MeV positron channelling}


\author*{\fnm{Matthew D.} \sur{Dickers\orcidlink{0000-0001-9615-9101}$^{\text{1*}}$}}\email{M.D.Dickers@kent.ac.uk}

\author{\fnm{Felipe} \sur{Fantuzzi\orcidlink{0000-0002-8200-8262}$^{\text{2}}$}}\email{f.fantuzzi@kent.ac.uk}

\author{\fnm{Nigel J.} \sur{Mason\orcidlink{0000-0002-4468-8324}$^{\text{1}}$}}\email{n.j.mason@kent.ac.uk}

\author{\fnm{Andrei V.} \sur{Korol\orcidlink{0000-0002-6807-5194}$^{\text{3}}$}}\email{korol@mbnexplorer.com}

\author{\fnm{Andrey V.} \sur{Solov'yov\orcidlink{0000-0003-1602-6144}$^{\text{3}}$}}\email{solovyov@mbnresearch.com}

\affil[1]{\orgdiv{Physics and Astronomy, School of Engineering, Mathematics and Physics}, \orgname{University of Kent}, \city{Canterbury}, \orgaddress{\street{Park Wood Rd}}, \postcode{CT2 7NH}, \country{United Kingdom}}

\affil[2]{\orgdiv{Chemistry and Forensic Science, School of Natural Sciences}, \orgname{University of Kent}, \city{Canterbury}, \orgaddress{\street{Park Wood Rd}}, \postcode{CT2 7NH}, \country{United Kingdom}}

\affil[3]{\orgname{MBN Research Center}, \orgaddress{\street{Altenh\"oferallee 3}}, \postcode{60438} \city{Frankfurt am Main}, \country{Germany}}


\abstract{We investigate the stability of crystalline undulator radiation (CUR) peaks emitted by \SI{530}{\mega\electronvolt} positron channelling in periodically bent C(110) crystals with varying bending amplitudes and bending periods. Relativistic molecular dynamics simulations were performed to quantify how these parameters affect the intensity and position of the CUR peak. The continuous potential approximation was used to identify isolines of constant peak energy, providing a reference for regions of spectral stability. MD results show that increasing the bending amplitude shifts the CUR peak to lower photon energies, while decreasing the period shifts it to higher energies, with both trends accompanied by enhanced dechannelling. For crystal parameters similar to recent experiments conducted at the MAinz MIkrotron (MAMI), the simulated CUR peak appears near \SI{0.515}{\mega\electronvolt}. These results demonstrate that the CUR peak remains stable across a broad range of bending amplitudes and periods, providing quantitative estimates of the sensitivity of the emitted radiation to variations in the crystal bending parameters.}


\maketitle

\section{Introduction} \label{sec:Intro}
Gamma-ray crystal-based light sources (CLS) present a novel approach to generating high-intensity electromagnetic radiation, comparable to and exceeding the energy ranges available at modern synchrotrons, X-ray free-electron lasers, and light sources based on laser Compton scattering \cite{Korol_2020}. Their operation is based on \textit{channelling}, the phenomenon in which charged particles (typically electrons or positrons) entering a crystal at small angles relative to the crystallographic planes or axes experience the collective influence of electrostatic fields of lattice atoms \cite{Lindhard_1965}. This leads to the transverse oscillation of the particles and the emission of \textit{channelling radiation} \cite{Kumakhov_1976}. Introducing a bend into the crystal structure leads to the emission of additional types of radiation: a uniform curvature leads to the emission of \textit{synchrotron-like radiation}. Periodic bending where the crystalline planes follow a sinusoidal profile lead to the emission of \textit{crystalline undulator radiation} (CUR) \cite{Korol_1998,KSG_review2004}, arising from the particle's undulation along the bent planes. Periodic bending is characterised by three key parameters: (i) the bending amplitude $a$, representing the maximum displacement of the crystalline planes from their unbent positions; (ii) the bending period $\lambda$, which defines the distance between successive extrema (maxima or minima); and (iii) the number of bending periods $N_{\text{u}}$. Adjusting these parameters allows control over the properties of the emitted radiation, such as the intensity and wavelength.

Several techniques have been proposed for manufacturing periodically
bent crystals, including mechanical bending \cite{Guidi_2011, Guidi_2007, Malagutti_2025}, etching, scratching, laser ablation and melting \cite{Bellucci_2003, Guidi_2005, Balling_2009, Bagli_2014}, sandblasting \cite{Camattari_2017}, ion implantation \cite{Bellucci_2015}, and crystal doping \cite{Breese_1997,Mikkelsen_2000,Krause_2002,Wistisen_EtAl:PRL_v112_254801_2014,ThuNhiTranThi_JApplCryst_v50_p561_2017}. Dynamical periodic bending by means of acoustic excitation suggested in Refs. \cite{BaryshevskyDubovskayaGrubich1980, Ikezi_1984, Dedkov_1994, Korol_1998, Korol_1999, Kaleris_2025} has been recently demonstrated experimentally \cite{HMU:JACA_v158_p4007_2025}.
While producing a crystal with the desired number of bending periods is relatively straightforward, reproducing the exact bending amplitude and bending period is far more difficult. Consequently, the exact bending amplitude and period of the crystal are often uncertain, leading to uncertainty in the properties of the emitted radiation. The sensitivity of the radiation characteristics to variations in the bending amplitude and period is well understood for tapered undulators used in free-electron lasers to maintain a consistent emission energy with changing electron beam characteristics \cite{Tomin_2024}, but has not been studied in detail for crystalline undulators.

Much work has been dedicated to the design and practical realisation of gamma-ray CLSs in recent years, and is the focus of the ongoing Horizon Europe EIC-Pathfinder-2021 Project TECHNO-CLS \cite{TECHNO-CLS}, where particular emphasis has been placed on studying CUR from both experimental and computational perspectives. Although many facilities offer high-quality electron beams, few offer positron beams. Positrons are generally more favourable for channelling because their positive charge confines them between the crystal planes. This reduces collisions with atomic nuclei that can lead to the loss of channelled particles, a process known as \textit{dechannelling}. After dechannelling, the particles undergo over-barrier motion, which in turn reduces the intensity of the emitted radiation. The MAinz MIkrotron (MAMI) accelerator facility recently began operating a \SI{530}{\mega\electronvolt} positron beamline \cite{Backe_2022}. The first channelling experiments using this beamline have recently been conducted \cite{Backe_2024, Mazzolari_2025}. Computational studies of the channelling process with electrons and positrons \cite{Pavlov_2020,Korol_2021, Sushko_2025, Korol_2025} have enabled estimates of CUR peak positions and intensities. However, CUR was only recently observed experimentally at MAMI \cite{Backe_2025} using \SI{855}{\mega\electronvolt} electron channelling in a quasi-periodically bent oriented diamond (110) crystal, C(110). Indications of CUR have been seen previously for \SI{270}{\mega\electronvolt} electrons in periodically bent Si(110) crystals with a bending period of $\lambda=\SI{9.9}{\micro\meter}$ \cite{Backe_2013}, as well as in Small-Amplitude Short-Period (SASP) Si crystals \cite{Wistisen_2014, Wistisen_2017}. In SASP crystals, the bending amplitude and period are much smaller than in a conventional crystalline undulator, such as those used in \citet{Backe_2025}; radiation emission arises from rapid, small-scale oscillations, rather than from the large-amplitude, undulator-like oscillations along the periodically bent profile of standard crystalline undulators \cite{Korol_2021, Korol2022Book}.

The goal of this study is to analyse the stability of the CUR peak position under variations in the crystal bending amplitudes and periods.
This stability can be thought of as a measure of how the CUR peak position varies when the bending amplitude and period are changed. Two approaches are employed: (i) an analytical estimation of the CUR peak position using the continuous potential approximation of \citet{Lindhard_1965}, which provides a quick estimate of the peak position; and (ii) relativistic molecular dynamics (MD) simulations \cite{Sushko_2013a}, offering a detailed description of the channelling dynamics and allowing the resulting radiation intensity to be calculated. While the dependence of CUR on the bending amplitude and period follows scaling laws prdicted by the continuous potential approximation, the quantitative sensitivity of the emitted radiation spectrum and intensity to realistic variations in these parameters has not previously been investigated using atomistic simulations. These simulations were performed using the \textsc{MBN Explorer} \cite{MBNExplorer} and \textsc{MBN Studio} \cite{MBNStudio} software packages, which enable atomistic-level investigation of particle dynamics and radiation emission. A comprehensive overview of the multiscale relativistic MD approach to modelling channelling and photon emission by ultra-relativistic charged particles alongside a number of case studies are presented in a recent review article \citet{Korol_2021}.

All analyses have been performed for \SI{530}{\mega\electronvolt} positron planar channelling in C(110) crystals, consistent with the parameters of the current MAMI positron beamline, assuming a periodically bent structure grown through chemical vapour deposition (CVD) and bent through boron doping \cite{Dickers_2025}. The results presented here help elucidate the stability of the CUR peak, quantify the sensitivity of the emitted radiation to variations in the bending amplitude and period, and identify ranges of bending parameters that produce radiation with well-defined spectral properties.

\section{Methodology} \label{sec:Method}
This section outlines the theoretical and computational methodologies used to analyse the stability of the CUR peak. The first part describes the continuous potential approximation, which is used to estimate the CUR peak positions over a range of bending amplitudes and periods, while the second part outlines the atomistic-level relativistic MD approach.

\subsection{Estimates from the Continuous Potential Approximation} \label{sec:CPA}
The channelling process can be described using the continuous (continuum) potential approximation \cite{Lindhard_1965}. In this model, ultra-relativistic charged particles that are incident at small angles with respect to the crystallographic planes experience collective effects from the crystal atoms. The particles move past many atoms over a very short time, therefore the discrete effect of individual atoms becomes `smeared' or averaged along the direction of motion, producing an effectively continuous inter-planar potential. Although this model neglects the discrete interactions between atoms and charged particles, it offers a useful analytical framework for estimating channelling conditions and the properties of the emitted radiation, and therefore can serve as a guide for more accurate relativistic MD simulations.

In a periodically bent crystal, a propagating particle experiences a centrifugal force as it follows the curved trajectory, with this force reaching a maximum at the extrema of the bending profile (i.e., at the maxima and minima). If this centrifugal force locally exceeds the maximum transverse restoring force $U'_{\text{max}}$ of the crystal's inter-planar potential, the particle may undergo over-barrier motion and dechannel. The influence of crystal bending is quantified by the parameter $C$, commonly known as the \textit{Tsyganov parameter} \cite{Tsyganov_1976a} or \textit{bending parameter}. It is defined as the ratio of the maximum centrifugal force experienced by the charged particle $F_{\text{cf}}^{\text{max}}$, which tends to drive it out of the channel, to the maximum transverse force confining it within the channel. For a periodically bent crystal this is written as \cite{KSG_review2004}:

\begin{equation}
    C = \frac{F_{\text{cf}}^{\text{max}}}{U'_{\text{max}}} = \frac{4\pi^2\varepsilon a}{\lambda^2U'_{\text{max}}}.
    \label{eqn:Bending_Parameter}
\end{equation}

\vspace{0.5 cm}

\noindent $\varepsilon$ is the particle energy (measured in \unit{\giga\electronvolt}), $\lambda$ is the period of the periodic bending, $a$ is its amplitude (both in \unit{\centi\meter}), and $U'_{\text{max}}\approx\SI{6}{\giga\electronvolt/\centi\meter}$ at \SI{300}{\kelvin} for C(110) \cite{Korol2014Book}. For a crystal with a weak curvature ($C\lesssim0.1$), the maximum centrifugal force is small relative to the inter-planar potential, and channelling is stable. For a strong curvature (\mbox{$C\gtrsim0.5$}), the maximum centrifugal force approaches or exceeds $U'_{\text{max}}$, increasing the probability of over-barrier motion and dechannelling. The parameter $C$ characterises the maximum centrifugal force experienced by a particle over one bending period. In reality, the local centrifugal force varies along the trajectory, ranging from zero along the centreline (where the curvature is zero), to its maximum value at the extrema. Thus, $C$ in \Cref{eqn:Bending_Parameter} provides a measure of the maximum curvature strength, while its local value defines the strength at a given point along the profile and thus the local stability of channelling.

As particles propagate through a periodically bent crystal they undergo two types of motion: (i) the undulation following the periodically bent profile, and (ii) oscillations within the inter-planar potential (between the crystalline planes for positrons, in the vicinity of the planes for electrons). The square of the average transverse velocity of a particle, i.e. the speed at which it undulates, may be quantified by the \textit{undulator parameter} $K$. Both types of motion contribute to the undulator parameter: \mbox{$K^2 = K_{\text{u}}^2 + K_{\text{ch}}^2(1-C)$} \cite{Korol_2001}, where the components for each type of undulation are given by:

\begin{gather}
    K_{\text{u}}^2 = \left(\frac{2\pi\gamma a}{\lambda}\right)^2 \label{eqn:Undulator_param_u}, \\
    K_{\text{ch}}^2 = \frac{2\gamma U_0}{3mc^2} \approx \num{2.55e-3}\varepsilon U_0 \label{eqn:Undulator_param_ch}.
\end{gather}

\vspace{0.5 cm}

The contribution from channelling oscillations, $K_{\text{ch}}^2$, is defined for positrons based on the harmonic approximation of the inter-planar potential in a straight channel \cite{Korol2014Book}. The factor $(1-C)$ is introduced to account for the curvature of the periodically bent crystal. Here, $U_0\approx\SI{20}{\electronvolt}$ is the depth of the potential well for C(110), and $\gamma=\varepsilon/mc^2$ is the Lorentz factor. For \SI{530}{\mega\electronvolt} positrons channelling in C(110), $K_{\text{ch}}^2\approx0.027$. The values of $K_{\text{u}}^2$ over the range of parameters considered in this study are outlined in Table S2 in the supplementary information.

For sufficiently large periods, the spectral distribution of emitted energy $\mathrm{d}E/\mathrm{d}(\hbar\omega)$ consists of two well separated peaks. The first corresponds to CUR, and is located at the position of the first harmonic. The second peak corresponds to channelling radiation. To understand the separation of these peaks, it is useful to estimate the wavelength of the channelling oscillations $\lambda_{\text{ch}}$, within the harmonic approximation of the inter-planar potential. This wavelength is given by:

\begin{equation}
    \lambda_{\text{ch}} \approx \pi d\sqrt{\frac{\varepsilon}{2U_0}},
\end{equation}

\vspace{0.5 cm}

\noindent where $d=\SI{1.2611}{\angstrom}$ is the (110) inter-planar distance in diamond \cite{Saotome_1998}, $U_0\approx\SI{20}{\electronvolt}$, as before, and $\varepsilon=\SI{530}{\mega\electronvolt}$. This gives a channelling oscillation wavelength of $\lambda_{\text{ch}}\approx\SI{1.44}{\micro\meter}$. Thus, for any values of the bending period much larger than the wavelength of the channelling oscillations, $\lambda\gg\lambda_{\text{ch}}$, the peaks from CUR and channelling radiation are well separated. This condition is satisfied for all values of $\lambda$ considered in this work, with the shortest bending period being $\lambda=\SI{3.5}{\micro\meter}>\lambda_{\text{ch}}$. We therefore restrict our discussion to the range of photon energies corresponding to the CUR peak. The position of this peak can be estimated from the first harmonic \cite{Elleaume_1992, Korol2014Book}:

\begin{equation}
    \hbar\omega_1 = \frac{9.5}{1+K^2/2}\frac{\varepsilon^2}{\lambda},
    \label{eqn:First_Harmonic}
\end{equation}

\vspace{0.5 cm}

\noindent where the factor of 9.5 originates from unit conversion, meaning
that $\varepsilon$ must be in \unit{\giga\electronvolt}, $\lambda$ in
\unit{\micro\meter}, and the value of $\hbar\omega_1$ is given in \unit{\mega\electronvolt} \cite{Korol2022Book}.
This equation can also be used to demonstrate that the CUR peak and channelling peak are well separated. For a periodically bent crystal C(110) crystal with a bending period of $\lambda=\SI{3.5}{\micro\meter}$ and amplitude $a=\SI{2.5}{\angstrom}$, the CUR peak is located at $\hbar\omega_1=\SI{0.686}{\mega\electronvolt}$. The position of the channelling peak can be found by substituting in \mbox{$\lambda=\lambda_{\text{ch}}=\SI{1.44}{\micro\meter}$}, giving $\hbar\omega=\SI{1.161}{\mega\electronvolt}$, thus for even the shortest bending period considered here, the two peaks will be well separated.

\begin{figure*}[t!]
    \centering
    \includegraphics[width=\textwidth]{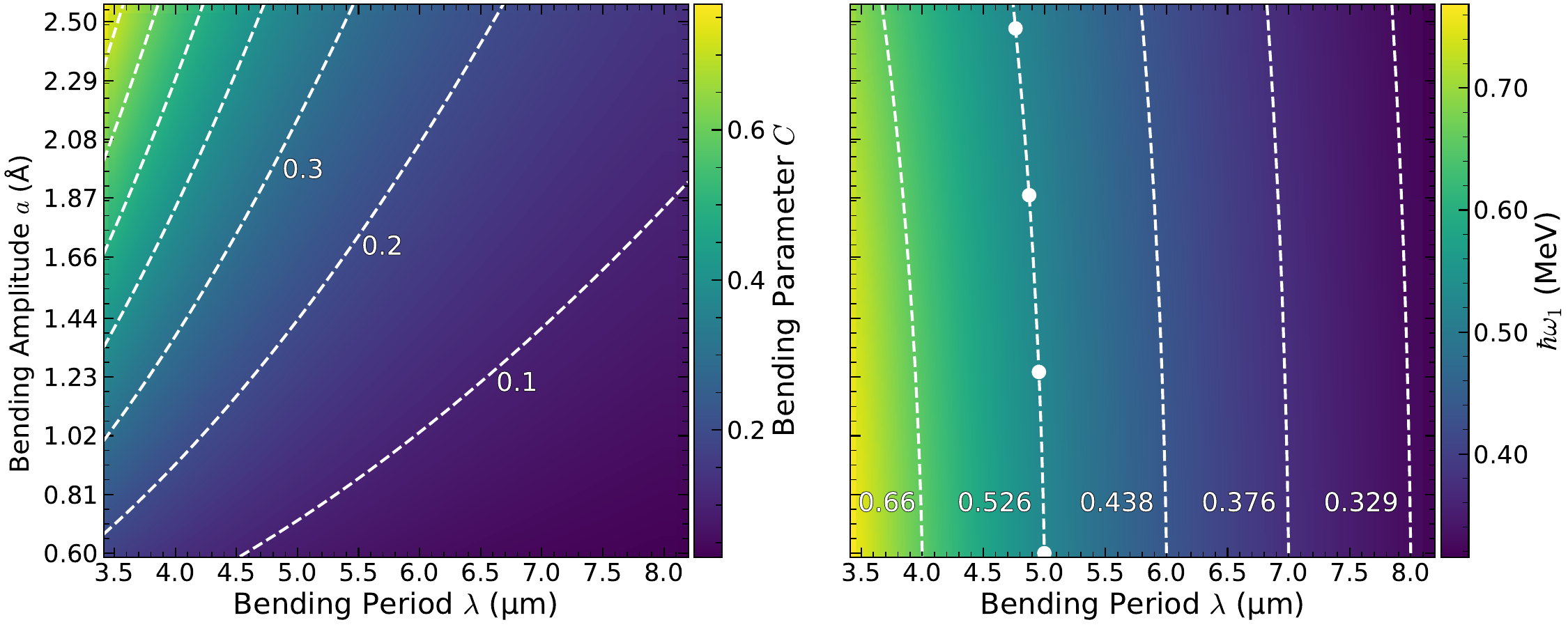}
    \caption{Heatmaps showing the variation in the values of the bending parameter $C$ \textbf{(left panel)} and first harmonic position $\hbar\omega_1$ \textbf{(right panel)} as a function of the bending period $\lambda$ and bending amplitude $a$. Both plots share the same $y$ axis. Each plot indicates various lines of constant $C$ and $\hbar\omega_1$ using dashed white lines with their corresponding values. Circular markers on the right panel indicate specific values along the isoline for $\hbar\omega_1=\SI{0.526}{\mega\electronvolt}$, discussed in detail later.}
    \label{fig:Heatmaps}
\end{figure*}

Within the continuous potential approximation, the position of the CUR peak
can be analysed as a function of the bending amplitude and bending period. Two key quantities describe this behaviour: the bending parameter $C$ and the first harmonic position $\hbar\omega_1$. When $C$ becomes too large, dechannelling dominates and the radiation intensity decreases. \Cref{fig:Heatmaps} shows the dependence of $C$ (left panel) and $\hbar\omega_1$ (right panel) on the bending amplitude and period. As expected, $C$ depends primarily on $a$, as a larger bending amplitude induces stronger centrifugal forces. In contrast, $\hbar\omega_1$ varies more strongly with $\lambda$: shorter bending periods and smaller amplitudes correspond to higher photon energies. The `isolines', indicated by dashed lines, highlight combinations of parameters that results in constant values of $C$ and $\hbar\omega_1$. For example, to maintain a photon energy of $\hbar\omega_1\approx\SI{0.526}{\mega\electronvolt}$, the bending period must remain within $\sim\SI{0.3}{\micro\meter}$ across the amplitude range of $a=0.60$ to \SI{2.50}{\angstrom}.

Although an approximation, the continuous potential model provides a reasonably good estimate of the effects of variable bending amplitude and period, and serves as an efficient bridge between experimental observations and atomistic simulations. However, it only provides an approximate description of the channelling efficiency through the bending parameter $C$, and an estimate of CUR peak positions through \Cref{eqn:First_Harmonic}, but no estimate of the radiation intensity. The continuous potential approximation therefore provides a useful means to estimate the viability of gamma-ray CLSs over a broad range of bending amplitudes and periods; however, a more detailed description of channelling dynamics and efficiency, as well as radiation intensity, can be obtained from atomistic simulations.

\subsection{Atomistic Modelling Approach}

Atomistic modelling provides a complementary description of channelling dynamics and radiation emission beyond the continuous potential approximation. Instead of averaging the inter-atomic potential across all atoms within the channel, the atomistic approach treats each atom as a discrete particle with its own interaction range. As a result, it explicitly accounts for stochastic interactions between propagating particles and crystal atoms. Alternative computational approaches, including Monte Carlo-based methods for particle propagation and radiation emission, have also been used to study the channelling phenomena in the context of CLSs \cite{Sytov_2019}. For a broader overview of available modelling approaches, see \citet{Korol_2021} and references therein.

Planar channelling simulations were conducted using the relativistic MD framework implemented in the \textsc{MBN Explorer} software package \cite{MBNExplorer}, developed for advanced multiscale modelling of molecular and nanoscale systems. The motion of ultra-relativistic charged particles within crystalline fields were modelled using the relativistic integrator for classical MD described in \citet{Sushko_2013a}, which also accounts for the changes in particle energy and velocity due to ionising collisions with crystal atoms \cite{Sushko_2025}.

To investigate the stability of the CUR peak, channelling simulations were conducted over a range of bending amplitudes and periods. The parameter space was selected in line with the crystal parameters used in recent experiments at MAMI \cite{Backe_2025}, corresponding to \mbox{$a\approx\SI{1.38}{\angstrom}$} and \mbox{$\lambda\approx\SI{5.0}{\micro\meter}$} with four undulator periods. A \mbox{$10\times10$} grid of amplitude-period combinations was selected to cover a broad range of $C$ values. The bending amplitudes spanned $a=0.60$ to \SI{2.50}{\angstrom} in steps of approximately \SI{0.21}{\angstrom}, and the bending periods from $\lambda=3.5$ to \SI{8.0}{\micro\meter} in steps of \SI{0.5}{\micro\meter}, as seen on the axes of \Cref{fig:Heatmaps}. The number of bending periods was fixed at ten, resulting in crystal lengths from 35 to \SI{80}{\micro\meter}. Although these crystals are longer than those used in \citet{Backe_2025}, this choice is compatible with potential future experiments with larger samples. For a direct comparison with existing results, the analysis can be restricted to the first four periods, as discussed later in \Cref{sec:Four_Period}.

For each amplitude–period pair, a periodically bent C(110) crystal structure was used, and channelling simulations using \SI{530}{\mega\electronvolt} positrons with zero divergence (i.e. perfectly collimated) were conducted. The process used to create these systems is detailed in \citet{Sushko_2013a}. In each case, \num{2000} independent positron trajectories were simulated using the Moli\'ere interaction potential \cite{Moliere_1947}. The initial particle entry coordinates were randomly distributed within a $3.78\times\SI{3.78}{\angstrom}$ (corresponding to three times the C(110) inter-planar distance $d=\SI{1.2611}{\angstrom}$) area at the crystal entrance to emulate the finite beam size in experiments, ensuring statistically robust results.

\section{Results and Discussion} \label{sec:R&D}
The results are presented in four sections. The first presents a statistical analysis of the channelling process, in which the trajectories of the propagating positrons are analysed. The second section presents an analysis of the spectral distribution of the emitted radiation. The third section analyses the stability of the CUR peak as predicted by the continuous potential model and the results of relativistic MD simulations. Finally, the fourth section discusses results of a four-period undulator, and provides an estimate of the CUR peak position in line with possible future positron channelling experiments at MAMI.

\subsection{Statistical Analysis of the Channelling Process} \label{sec:Trajectory_Analysis}
Due to the randomised entry locations of the particles at the crystal entrance, not all particles begin propagating in the channelling mode; those that do are known as \textit{accepted} particles. The efficiency of this process can be quantified by the acceptance $\mathcal{A}=N_{\text{acc}}/N_{\text{0}}$, the ratio of the number of accepted particles $N_{\text{acc}}$ to the total number of incident particles $N_0$. A table of acceptance values covering the range of bending amplitudes and periods considered is provided in Table S5 in the supplementary information.

The number of particles in the channelling mode at any penetration distance $z$ through the crystal is influenced by dechannelling and rechannelling (the re-entry of particles into the channelling mode) events. These processes are reflected in the channelling fractions \mbox{$f_{\text{ch},0}(z) = N_{\text{ch},0}(z)/N_{\text{acc}}$} and \mbox{$f_{\text{ch}}(z) = N_{\text{ch}}(z)/N_{\text{acc}}$} \cite{Sushko_2013a, Korol_2021}. $N_{\text{ch},0}(z)$ is the number of accepted particles that remain in the channelling mode at a given penetration distance, and $N_{\text{ch}}(z)$ is the total number of channelled particles, including those that have rechannelled. Both fractions are normalised to the number of accepted particles, $N_{\text{acc}}$, so the number of particles at any distance $z$ can be recovered by multiplying by $N_{\text{acc}}$. It is important to note that this normalisation may lead to values of $f_{\text{ch}}(z)$ exceeding one, as particles that were not initially accepted may subsequently enter the channelling mode through rechannelling, resulting in $N_{\text{ch}}(z) > N_{\text{acc}}$.

\begin{figure*}[t!]
    \centering
    \includegraphics[width=\textwidth]{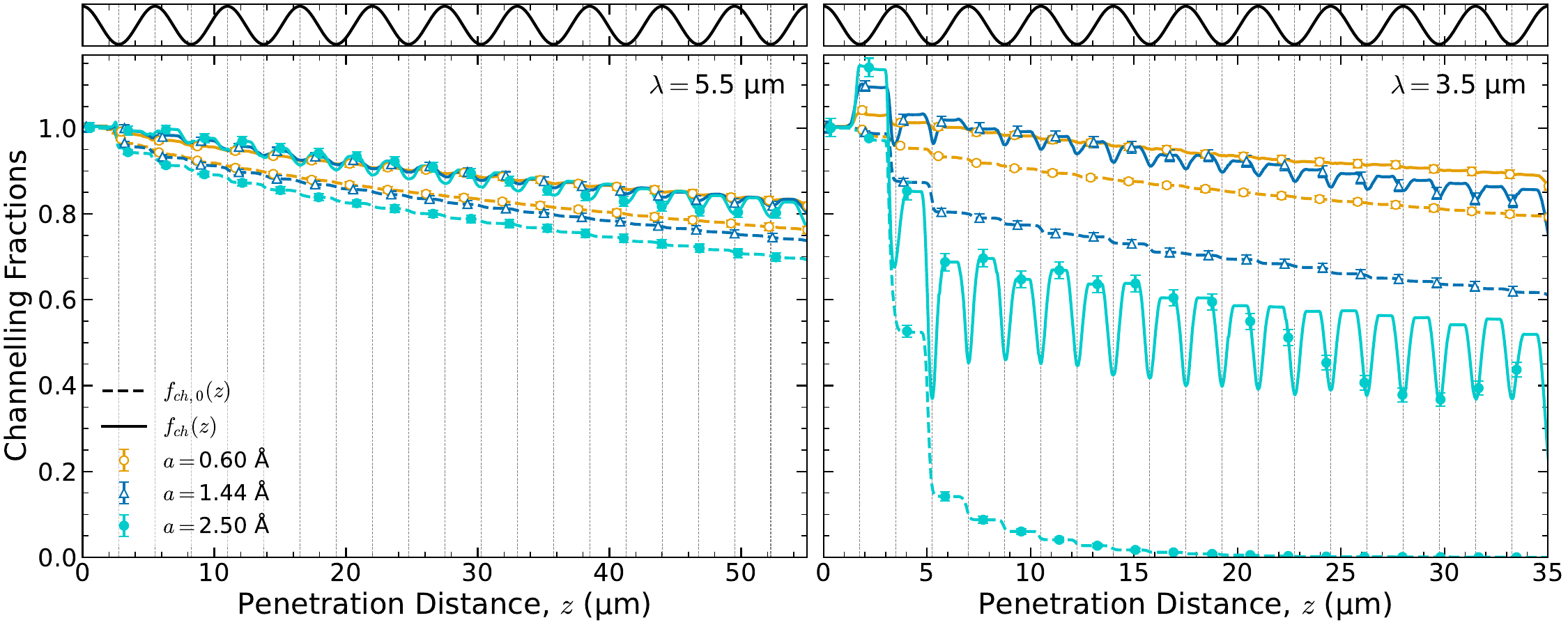}
    \caption{Channelling fractions $f_{\text{ch},0}$ (dashed lines) and $f_{\text{ch}}$ (solid lines) as a function of penetration distance for bending periods of $\lambda=\SI{5.5}{\micro\meter}$ \textbf{(left panel)} and \SI{3.5}{\micro\meter} \textbf{(right panel)}. Both plots show bending amplitudes of $a=0.60$, 1.44, and \SI{2.50}{\angstrom}. A representation of the bending profile is included at the top of each plot, with vertical lines corresponding to the maxima and minima of the profile. Note that the fractions are normalised to the number of accepted particles $N_{\text{acc}}$.}
    \label{fig:Channelling_Fractions}
\end{figure*}

\Cref{fig:Channelling_Fractions} shows these fractions for the smallest, intermediate, and largest bending amplitudes, $a = 0.60$, 1.44, and \SI{2.50}{\angstrom}, and for bending periods
$\lambda = 5.5$ and \SI{3.5}{\micro\meter}. Increasing the bending amplitude leads to increased dechannelling, shown by the steeper decrease of both $f_{\text{ch},0}(z)$ and $f_{\text{ch}}(z)$. This behaviour follows directly from the dependence of the bending parameter $C$ on the bending amplitude $a$, \Cref{eqn:Bending_Parameter}; a larger $a$ increases the maximum centrifugal force, and thus the probability of dechannelling. The same behaviour is seen in both fractions, though $f_{\text{ch}}(z)$ also includes the effects of rechannelling.

The fluctuations of $f_{\text{ch}}(z)$, characterised by periodic decreases and recoveries, reflect the competition between dechannelling and rechannelling events \cite{Korol_2017}. While the bending parameter $C$ defined in \Cref{eqn:Bending_Parameter} is the maximum value characterising the overall bending parameter, the local centrifugal force experienced by a particle varies along the trajectory. This force reaches a maximum at the extrema of the bending profile, causing a significant fraction of particles to leave the channelling mode, which corresponds to dechannelling and produces the minima in $f_{\text{ch}}(z)$. Conversely, at the midpoints of the bending profile, where the local curvature, and therefore the centrifugal force is zero, the bending profile is effectively quasi-linear, and particles are more likely to be recaptured into the channelling mode, resulting in a recovery of $f_{\text{ch}}(z)$: rechannelling. In this way, the observed periodic modulation of $f_{\text{ch}}(z)$ is directly linked to the variation in the local centrifugal force along the bending profile.

This pattern is observed for both bending periods, although it is more pronounced for the shorter period $\lambda = \SI{3.5}{\micro\meter}$. Since $C\propto a/\lambda^2$, shorter periods correspond to larger curvature and stronger centrifugal forces, which enhance dechannelling. Increasing the amplitude further amplifies this effect, resulting in lower overall channelling fractions. This explains the increases of $f_{\text{ch}}(z)$ during the first half of the first period, where the normalised channelling fraction exceeds one. In this region, particles that were not initially accepted at the crystal entrance are captured into the channelling mode within the quasi-linear segment of the bending profile. Upon reaching the bending minima at $z=\lambda/2$, most of these captured particles dechannel and undergo over-barrier motion.

An interesting feature emerges when comparing $f_{\text{ch}}(z)$ for $a = 0.60$ and \SI{1.44}{\angstrom}. In the first few periods, the crystal with the largest bending amplitude exhibits a slightly higher value of $f_{\text{ch}}(z)$, contrary to what one might expect from \Cref{eqn:Bending_Parameter}. With increasing penetration distance the two curves converge, beyond which $f_{\text{ch}}(z)$ for \mbox{$a = \SI{0.60}{\angstrom}$} becomes greater, as expected. This effect is observed for both $\lambda = 5.5$ and \SI{3.5}{\micro\meter}, but is more pronounced for the shorter period. It should be noted that the differences are slightly exaggerated due to the normalisation by $N_{\text{acc}}$. For $a=\SI{1.44}{\angstrom}$, the values of $\mathcal{A}$ for $\lambda=5.5$ and \SI{3.5}{\micro\meter} are 0.868 and 0.678 respectively. The initial enhancement of $f_{\text{ch}}(z)$ for larger amplitudes arises from the longer length of the quasi-linear segment compared to smaller amplitudes, allowing for greater rechannelling. However, since the maximum bending parameter $C$ also increases with amplitude, dechannelling dominates over the full crystal length, leading to a faster long-term decay of $f_{\text{ch}}(z)$. In contrast, smaller amplitudes have shorter quasi-linear regions, which suppress rechannelling and result in a smoother overall decay of $f_{\text{ch}}(z)$.

This interplay also explains why the fluctuations of $f_{\text{ch}}(z)$ are less pronounced for smaller bending amplitudes. For small $a$, the local curvature, and thus the variation in the centrifugal force are relatively small, leading to less pronounced dechannelling at the extrema and a smoother overall decay of $f_{\text{ch}}(z)$. In contrast, larger amplitudes result in larger variations in centrifugal force, enhancing dechannelling at the extrema and increasing rechannelling in the quasi-linear regions, resulting in larger fluctuations in $f_{\text{ch}}(z)$.


\begin{figure*}[t!]
    \centering
    \includegraphics[width=\textwidth]{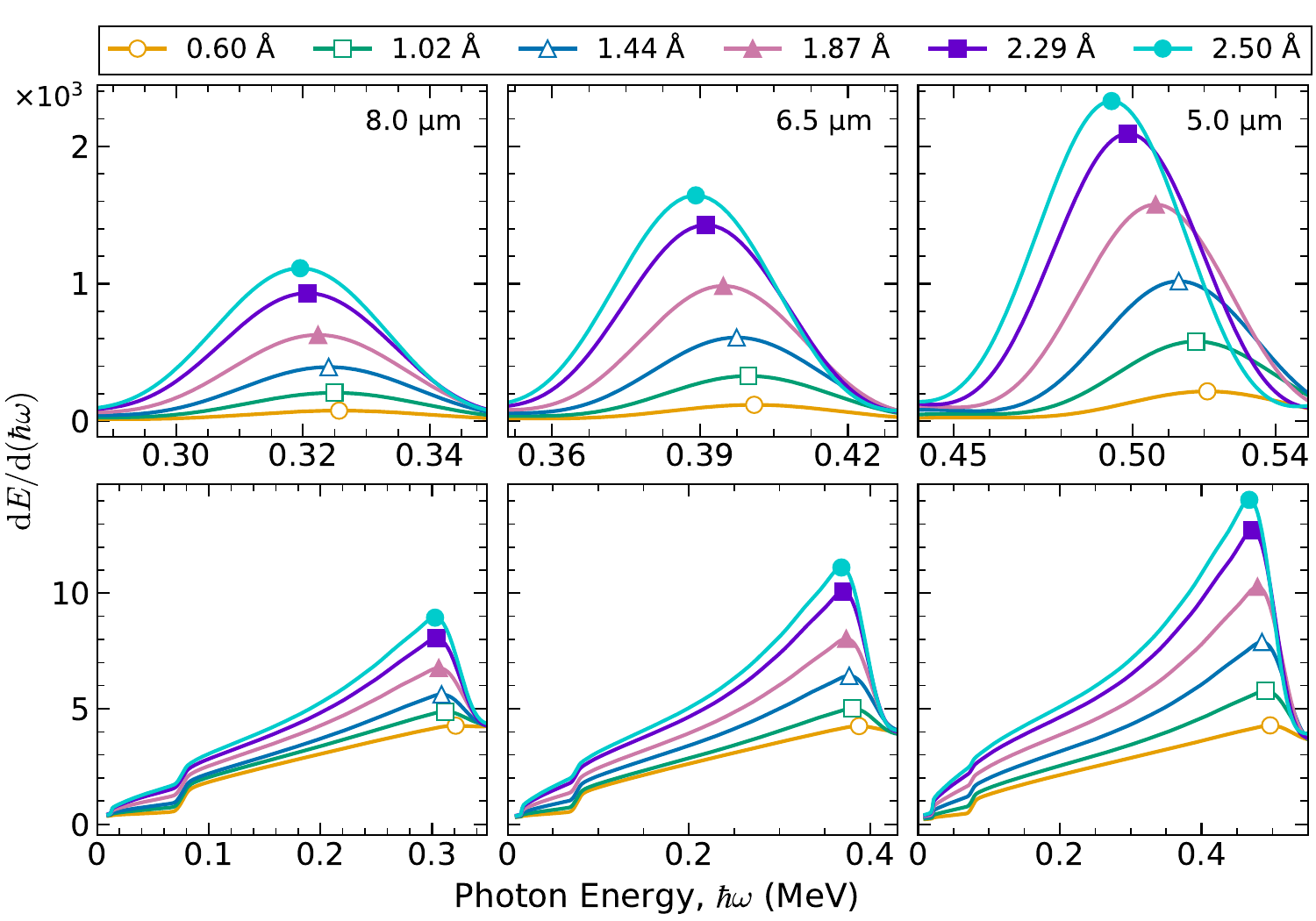}
    \caption{Spectral distributions of crystalline undulator radiation for \SI{530}{\mega\electronvolt} positrons for different bending profiles. The top row corresponds to the emission cone $\theta_0=\SI{133.8}{\micro\radian}\approx0.139/\gamma$, and the bottom row $5/\gamma \approx \SI{4821}{\micro\radian}$. The left column corresponds to a bending period of $\lambda=\SI{8.0}{\micro\meter}$, the middle column \SI{6.5}{\micro\meter}, and the right column \SI{5.0}{\micro\meter}. All cases show the same range of bending amplitudes $a$, indicated by the top legend. In all cases markers indicate the spectral peaks.}
    \label{fig:Spectra_All}
\end{figure*}

\subsection{Spectral Distribution of Undulator Radiation} \label{sec:Spectra_Analysis}

The spectral angular distribution $\mathrm{d}^3E/\mathrm{d}(\hbar\omega)\mathrm{d}\Omega$ of the emitted energy $\hbar\omega$ within the solid angle $\mathrm{d}\Omega\approx\theta \mathrm{d}\theta \mathrm{d}\phi$ was calculated based on the recorded particle trajectories. The numerical algorithm used is detailed in Refs.~\cite{Sushko_2013a, Korol2014Book}, and is implemented within \textsc{MBN Explorer}. The spectral distribution of the radiation emitted within a cone $\theta_0 \ll 1$ may be calculated along any given direction, averaged over all $N_0$ simulated trajectories using the following equation:
\begin{equation}
    \frac{\mathrm{d}E(\theta \leq \theta_0)}{\mathrm{d}(\hbar\omega)} = \frac{1}{N_0}\sum_{n=1}^{N_0}\int_0^{2\pi}\mathrm{d}\phi\int_0^{\theta_0}\theta \mathrm{d}\theta \frac{\mathrm{d}^3E_n}{\mathrm{d}(\hbar\omega)\mathrm{d}\Omega},
    \label{eqn:Spectral_Distribution}
\end{equation}
\vspace{0.5 cm}

\noindent which accounts for contributions from all trajectory types. In this work, we focus exclusively on the stability of the CUR peak, and do not analyse the peak corresponding to channelling radiation. As discussed in \Cref{sec:CPA}, for \SI{530}{\mega\electronvolt} positrons the CUR and channelling radiation peaks are well separated, with the channelling peak appearing at much higher photon energies ($\gtrsim\SI{1}{\mega\electronvolt}$) \cite{Sushko_2025}. This channelling radiation still contributes a smooth background at lower photon energies. In the following analyses, only photon energies in the vicinity of the CUR peak are considered. For further relativistic MD simulations considering channelling radiation in linear crystals, see \citet{Sushko_2025}.

\begin{figure*}[t!]
    \centering
    \includegraphics[width=\textwidth]{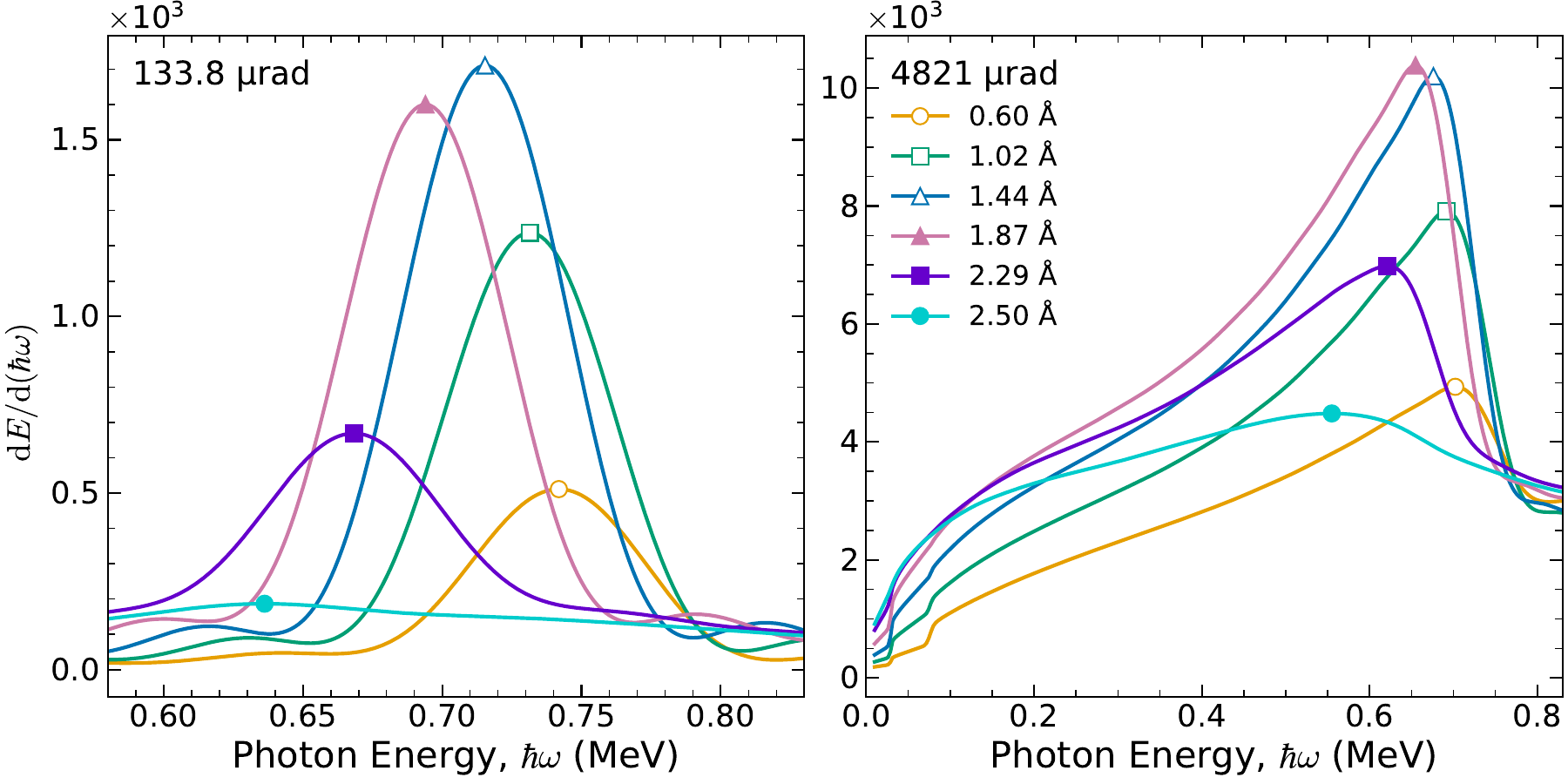}
    \caption{Spectral distributions of crystalline undulator radiation for \SI{530}{\mega\electronvolt} positrons for the emission cones $\theta_0=\SI{133.8}{\micro\radian}\approx0.139/\gamma$ \textbf{(left panel)} and $\theta_0=5/\gamma\approx\SI{4821}{\micro\radian}$ \textbf{(right panel)}. Both cases are for a bending period of $\lambda=\SI{3.5}{\micro\meter}$. In all cases markers indicate the peak spectral peaks.}
    \label{fig:Spectra_3.5}
\end{figure*}

\Cref{fig:Spectra_All} shows the spectral distributions of CUR for two emission cones along the undulator axis: $\theta_0 = \SI{133.8}{\micro\radian}\approx0.139/\gamma$ (top row), corresponding to the beam aperture at MAMI \cite{Backe_2025}; and $\theta_0 = 5/\gamma \approx \SI{4821}{\micro\radian}$ (bottom row), encompassing the full range of emitted radiation. For reference, the natural emission cone for \SI{530}{\mega\electronvolt} positrons is $\theta_0=1/\gamma\approx\SI{964}{\micro\radian}$. Each column corresponds to a fixed bending period, namely $\lambda = 8.0$, 6.5, and \SI{5.0}{\micro\meter}, and shows spectra for different bending amplitudes $a$. Across both emission cones, similar trends are observed: decreasing the bending period increases the spectral intensity and shifts the CUR peak to higher photon energies. Conversely, increasing the bending amplitude increases the spectral intensity but shifts the peak to lower photon energies. This behaviour reflects the dependence of the first harmonic energy $\hbar\omega_1$ (\Cref{eqn:First_Harmonic}) on the bending amplitude and period. The intensity of radiation emitted from a charged particle is proportional to the square of its acceleration \cite{Jackson_1998}. Therefore, shorter bending periods and larger bending amplitudes increase the transverse acceleration of the particle, enhancing the intensity of the emitted radiation.

However, at the shortest bending period, $\lambda = \SI{3.5}{\micro\meter}$, the behaviour differs noticeably, acquiring a non-monotonic dependence on the bending amplitude, as shown in \Cref{fig:Spectra_3.5}. For both emission cones, the spectral intensity increases with amplitude up to between $a=1.44$ and \SI{1.87}{\angstrom}, but then decreases for larger amplitudes, although the CUR peak positions remain consistent with the observations in \Cref{fig:Spectra_All}. This reduction correlates with enhanced dechannelling, as discussed in \Cref{sec:Trajectory_Analysis} and illustrated in \Cref{fig:Heatmaps}. For $\lambda = \SI{3.5}{\micro\meter}$, when $a \gtrsim \SI{1.87}{\angstrom}$, the bending parameter exceeds $C > 0.5$, leading to significant dechannelling and thus a decrease in the spectral intensity. At $a \approx \SI{2.5}{\angstrom}$, where $C \gtrsim 0.7$, dechannelling becomes dominant, strongly suppressing the spectral intensity.

\begin{figure*}[t!]
    \centering
    \includegraphics[width=\textwidth]{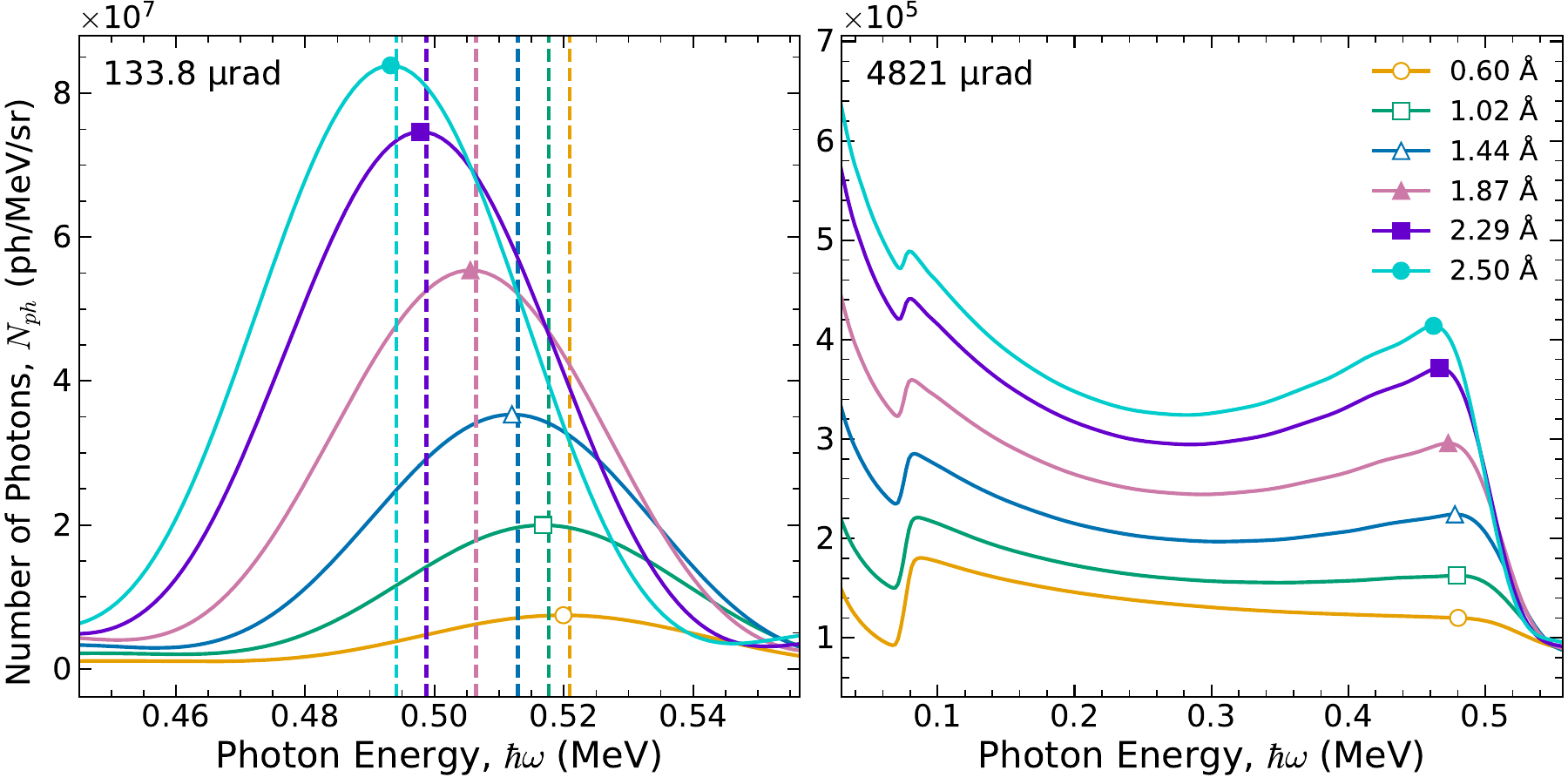}
    \caption{Number of photons in units of photons per \unit{\mega\electronvolt} per solid angle $\Delta\Omega \approx\pi\theta_0^2$ for the emission cones $\theta_0=\SI{133.8}{\micro\radian}\approx0.139/\gamma$ \textbf{(left panel)} and $\theta_0=5/\gamma\approx\SI{4821}{\micro\radian}$ \textbf{(right panel)}. Both cases are for a bending period of $\lambda=\SI{5.0}{\micro\meter}$. In all cases markers indicate the peak number of photons. The dashed lines in the left panel indicate the spectral peaks from \Cref{fig:Spectra_All}.}
    \label{fig:Num_Photons}
\end{figure*}

Although the continuous potential approximation provides reasonable estimates of the CUR peak position over much of the parameter space considered here, its applicability becomes increasingly limited for large values of $C$. In this regime, enhanced dechannelling leads to a significant reduction in the channelling fraction, as discussed in \Cref{sec:Trajectory_Analysis}, and thus suppresses the radiation intensity. At these larger $C$ values, the continuous potential approximation should therefore be treated as providing qualitative guidance, whereas the atomistic simulations retain a quantitative description of both the channelling dynamics and the resulting radiation properties. Nevertheless, the good agreement observed between the two approaches over the parameter range considered demonstrates that, even at relatively large values of $C$, the continuous potential approximation remains a useful tool for estimating viable bending parameters and the corresponding CUR peak positions for the design of gamma-ray CLSs.

It is important to note that the results presented here assume a perfectly collimated positron beam. In reality, the positron beam at MAMI has a  horizontal and vertical angular divergence of approximately $640\times\SI{64}{\micro\radian}$ \cite{Backe_2022}. This will lead to some minor broadening and reduction in the intensity of the spectral distributions presented in \Cref{fig:Spectra_All,fig:Spectra_3.5}. Importantly, it is unlikely to have a significant impact on the position of the spectral peaks. A larger beam divergence results in particles entering the crystal with a wider range of incident angles. Accepted particles may therefore exhibit larger channelling amplitudes, increasing the contribution of $K^2_{\text{ch}}$ to the total undulator parameter. However, for the relatively small positron beam energies of \SI{530}{\mega\electronvolt} used at MAMI in diamond, $K^2\ll1$, so the resulting shift in the CUR peak position is expected to be very small.

The spectral distribution of the number of photons within a unit emission cone can be calculated from  \Cref{eqn:Spectral_Distribution} by dividing by $\hbar\omega$ and the solid angle $\Delta\Omega\approx\pi\theta_0^2$. \Cref{fig:Num_Photons} shows the resulting photon emission spectra for $\lambda=\SI{5.0}{\micro\meter}$, corresponding to the parameters used in the MAMI experiment \cite{Backe_2025}. As with the spectral intensity, increasing the bending amplitude results in a higher number of photons. Compared with the spectral distributions shown in \Cref{fig:Spectra_All,fig:Spectra_3.5}, the peak number of photons is shifted to slightly lower energies, as indicated by the dashed lines marking the peak positions from \Cref{fig:Spectra_All}.

A small peak is seen at photon energies of approximately \SI{0.1}{\mega\electronvolt} in the bottom right panel of \Cref{fig:Spectra_All}. This feature arises from the truncation of the radiation spectrum at lower photon energies due to the finite emission cone $\theta_0=\SI{4821}{\micro\radian}$. For a larger (or infinite) emission cone, the spectral intensity would extend smoothly without the observed peak. The effect becomes more pronounced in the right panel of \Cref{fig:Num_Photons}, since dividing by $\hbar\omega$ amplifies contributions at low photon energies.

\subsection{Stability of the Crystalline Undulator Radiation Peak}

\begin{figure}[t!]
    \centering
    \includegraphics[width=\columnwidth]{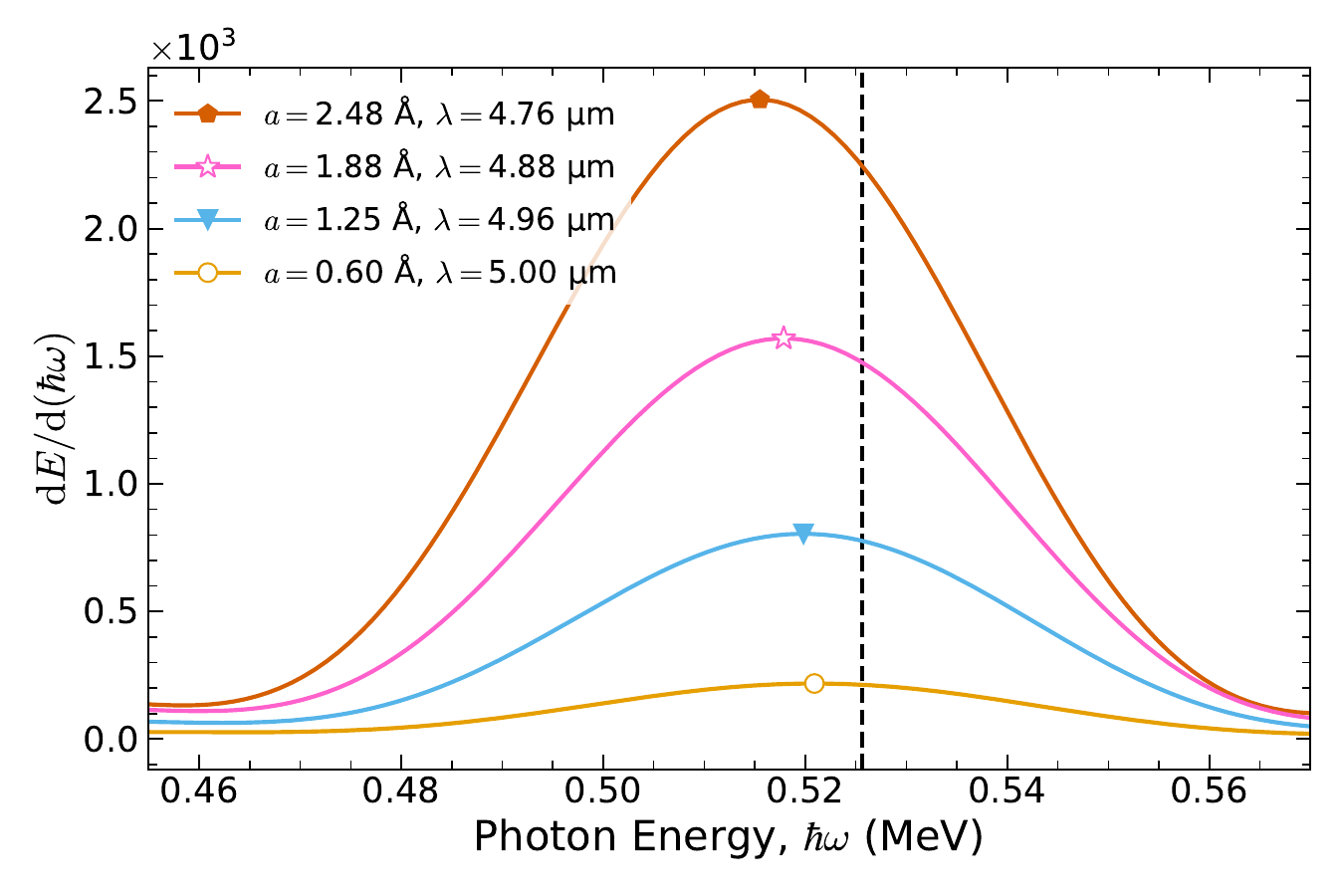}
    \caption{Spectral distributions of crystalline undulator radiation for \SI{530}{\mega\electronvolt} positrons for the emission cone $\theta_0=\SI{133.8}{\micro\radian}\approx0.139/\gamma$ with bending amplitudes and periods along the isoline for $\hbar\omega_1\approx\SI{0.526}{\mega\electronvolt}$ from the right panel of \Cref{fig:Heatmaps}. The peak positions are indicated by markers. The black dashed line indicates the value of $\hbar\omega_1$ as calculated with \Cref{eqn:First_Harmonic}.}
    \label{fig:Iso_Plots}
\end{figure}

In \Cref{fig:Heatmaps}, isolines of constant $\hbar\omega_1$ were estimated by means of the continuous potential approximation for various combinations of bending amplitude and period. To validate these results four points along the isoline for $\hbar\omega\approx\SI{0.526}{\mega\electronvolt}$ were selected, and full atomistic channelling simulations were performed as previously described. These points were chosen approximately equidistant along the isoline and can be seen in \Cref{fig:Heatmaps}: $a=\SI{2.48}{\angstrom}$ and $\lambda=\SI{4.76}{\micro\meter}$, $a=\SI{1.88}{\angstrom}$ and $\lambda=\SI{4.88}{\micro\meter}$, $a=\SI{1.25}{\angstrom}$ and $\lambda=\SI{4.96}{\micro\meter}$, and $a=\SI{0.60}{\angstrom}$ and $\lambda=\SI{5.00}{\micro\meter}$. \Cref{fig:Iso_Plots} shows the resulting spectral distributions of CUR, with the predicted peak position from \Cref{eqn:First_Harmonic} indicated as a black dashed line. A slight shift to lower photon energies is observed as the bending amplitudes increase and the bending period decreases, consistent with the expected dependence on $a$ and $\lambda$. However, the total shift is only \SI{5.4e-3}{\mega\electronvolt} between the smallest and largest amplitudes. Such a small variation may be difficult to resolve experimentally due to the finite resolution of detector systems. Nevertheless, this confirms that the position of the CUR peak does not change noticeably when evaluated using the atomistic approach and validates the use of the continuous potential approximation for estimating isolines of constant $\hbar\omega_1$ on \Cref{fig:Heatmaps}.

It should be noted that the simulated peak positions lie at slightly smaller photon energies compared to the theoretical value predicted by \Cref{eqn:First_Harmonic}. This difference arises from the finite emission cone used here ($\theta_0=\SI{133.8}{\micro\radian}$), whereas \Cref{eqn:First_Harmonic} assumes emission only in the forward direction, $\theta_0=0$. As shown in \Cref{fig:Spectra_All}, increasing the size of the emission cone shifts the CUR peak to lower photon energies. Therefore, for $\theta_0=0$, the estimated CUR peak will be at higher photon energies. These results verify that any point chosen along the isolines of $\hbar\omega_1$ in \Cref{fig:Heatmaps} provides a reliable estimate for the position of the CUR peak.

The same peak behaviour is evident in the results presented in \Cref{fig:Spectra_All}. The separation between CUR peaks across different bending amplitudes for a fixed bending period decreases as the bending period decreases: for $\lambda=\SI{8.0}{\micro\meter}$ the change in CUR peak position between the smallest and largest bending amplitudes is \SI{6.2e-3}{\mega\electronvolt}, while for $\lambda=\SI{5.0}{\micro\meter}$ the shift is \SI{0.02}{\mega\electronvolt}.

\subsection{Four-Period Undulator Comparison} \label{sec:Four_Period}

\begin{figure}[t!]
    \centering
    \includegraphics[width=\columnwidth]{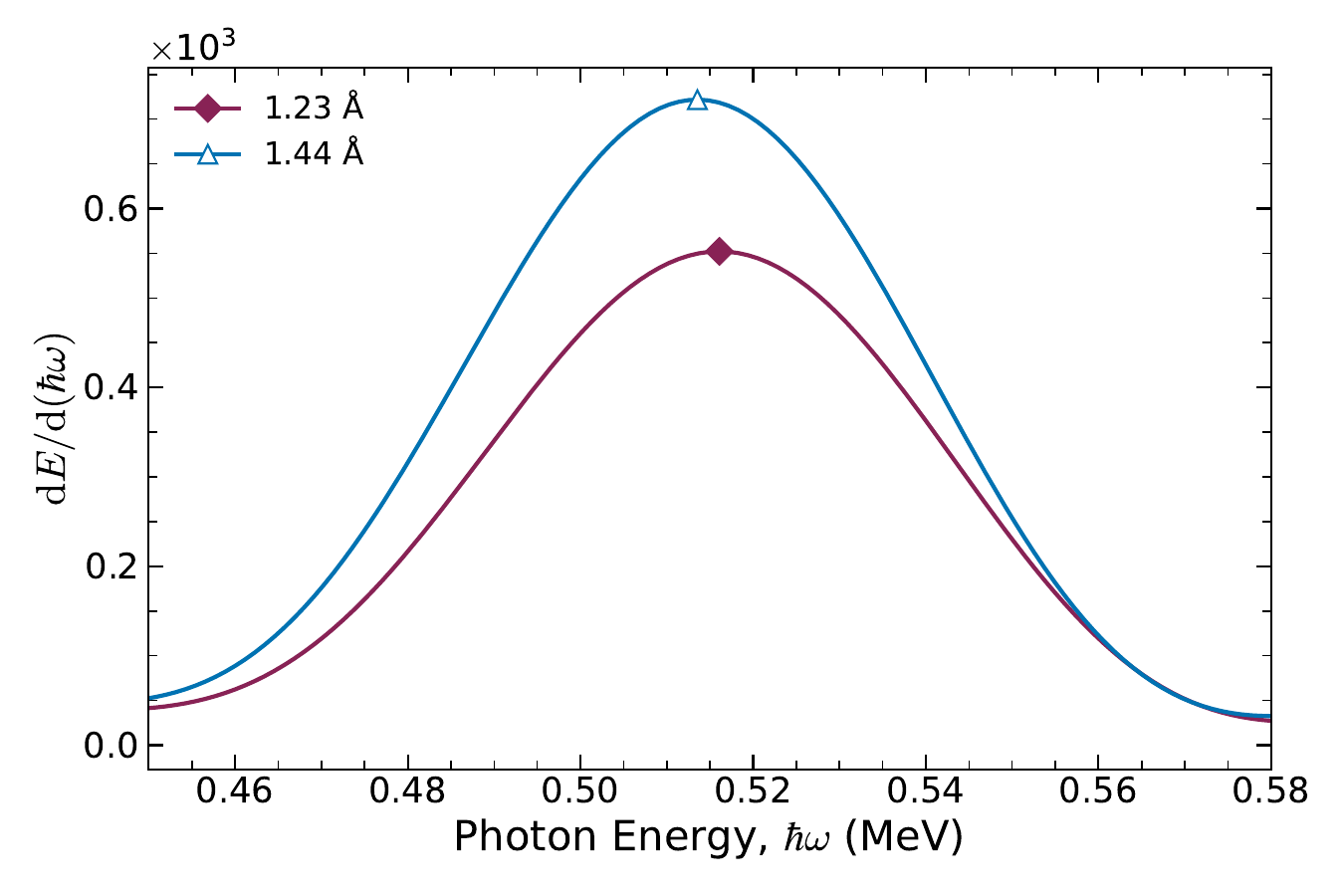}
    \caption{Spectral distributions of crystalline undulator radiation for \SI{530}{\mega\electronvolt} positrons for the emission cone $\theta_0=\SI{133.8}{\micro\radian}\approx0.139/\gamma$ for a four-period undulator with a bending period of $\lambda=\SI{5.0}{\micro\meter}$ (equivalent to that of MAMI) and two bending amplitudes $a=1.23$ and \SI{1.44}{\angstrom}. The peak positions are indicated by markers.}
    \label{fig:Four_Periods}
\end{figure}

The recent \SI{855}{\mega\electronvolt} electron channelling experiments conducted at MAMI \cite{Backe_2025} employed C(110) crystals with four bending periods, in contrast to the ten periods considered in this work. For a direct comparison, spectral distributions were recalculated over only the first four undulator periods for two crystals closely matching the MAMI setup: $\lambda=\SI{5.0}{\micro\meter}$ with $a=1.23$ and \SI{1.44}{\angstrom}. These amplitudes are close to the experimental value $a\approx\SI{1.38}{\angstrom}$ used at MAMI \cite{Backe_2025}. The spectral distributions were calculated for the emission cone corresponding to the aperture at MAMI, $\theta_0=\SI{133.8}{\micro\radian}$, and are shown in \Cref{fig:Four_Periods}.

The most notable difference compared to the ten-period case discussed
previously in \Cref{sec:Spectra_Analysis} is an approximately 1.4-fold
decrease in the spectral intensity (see the top-right graph in \Cref{fig:Spectra_All}). For on-axis emission in the forward direction ($\theta_0=0$), the spectral intensity $I$ scales with the square of the number of bending periods $I\propto N_{\text{u}}^2$, while for emission over a finite cone ($\theta_0>0$) along the same axis, the scaling is linear, $I\propto N_{\text{u}}$ \cite{Howells_1994}. Thus, for an ideal harmonic crystalline undulator, the intensity should increase by a factor of $10/4=2.5$ when the number of periods is increased from 4 to 10, significantly larger than the factor of 1.4 observed from simulations.

Two effects can account for this discrepancy. First, the linear scaling of intensity with the number of bending periods assumes that particles channel through the full length of the crystal; however, dechannelling reduces the number of particles in the channelling mode and thus their contribution to emission. From \Cref{fig:Channelling_Fractions}, the channelling fractions for 4 and 10 periods are approximately $f_{\text{ch}}(N_{\text{u}}=4)=0.9$ and $f_{\text{ch}}(N_{\text{u}}=10)=0.8$, which reduces the intensity ratio to $2.2$, still well above the simulated value.

The second, more significant, factor arises from the finite size of the emission cone $\theta_0$. The cone corresponding to the setup at MAMI is \mbox{$\theta_0=\SI{133.8}{\micro\radian}$} \cite{Backe_2025}, which is approximately 7.2 times smaller than the natural emission cone for \mbox{$\theta_0=1/\gamma\approx\SI{964}{\micro\radian}$} for \SI{530}{\mega\electronvolt} positrons. Therefore, the measured intensity is only integrated over a small fraction of the total emitted radiation. In a crystalline undulator, the angular distribution of radiation from successive bending periods narrows with an increasing number of periods. The natural angular width $\Theta_{N_{\text{u}}}$ is given by $\Theta_{N_{\text{u}}} \approx 1/\gamma\sqrt{N_{\text{u}}}$ \cite{Schmuser_2009}. Within this cone, the energy of the harmonic remains approximately stable. The ratio of the sizes of these emission cones $\Theta_{10}/\Theta_4 \approx \sqrt{4/10} \approx 0.63$ indicates that the emission cone decreases by 37\%. Since the MAMI emission cone is smaller than the natural emission cone, this narrowing reduces the fraction of radiation collected by around the same factor. Accounting for reductions in intensity from both dechannelling and narrowing emission cones, the reduction in intensity between 10 and 4 periods is $2.5\times0.89\times0.63\approx1.4$, consistent with the observed reduction in intensity from the simulation results.

These results provide an estimate of the CUR peak position for bending amplitudes and periods very close to those used in the recent electron channelling experiments \cite{Backe_2025}, and possible future experiments with the \SI{530}{\mega\electronvolt} positron beamline and detector setup. Taking the average CUR peak position across the two simulated cases (corresponding to a bending amplitude of $a=1.34$ close to the experimental $\sim\SI{1.38}{\angstrom}$), the CUR peak should be located at a photon energy of approximately \SI{0.515}{\mega\electronvolt}. This estimate may assist in refining the photon energy search range for identifying the CUR peak in future positron channelling experiments.

\section{Conclusions and Outlook} \label{sec:Conclusions}

In this paper we have investigated how the intensity and position of the CUR peak for \SI{530}{\mega\electronvolt} positrons channelling in periodically bent oriented diamond (110) crystals, C(110), depend on the crystal bending amplitude $a$ and period $\lambda$. Relativistic MD simulations were used to model positron channelling across a range of bending amplitudes ($a=0.60-\SI{2.50}{\angstrom}$) and bending periods ($\lambda=3.5-\SI{8.0}{\micro\meter}$), using crystal parameters employed in recent experiments conducted at MAMI \cite{Backe_2025}.

These simulations showed that increasing the bending amplitude shifts the CUR peak to lower photon energies, while decreasing the bending period shifts it to higher photon energies. The spectral intensity generally increases with larger amplitudes or shorter periods. For the shortest bending period ($\lambda=\SI{3.5}{\micro\meter}$), the spectral intensity decreased beyond $a=\SI{1.44}{\angstrom}$, which is attributed to increased dechannelling. This was corroborated by statistical analysis of positron trajectories. Along isolines corresponding to a fixed photon energy $\hbar\omega_1\approx\SI{0.526}{\mega\electronvolt}$, the CUR peak position varies by only \SI{5.4e-3}{\mega\electronvolt}. This also allows for the identification of sets of parameters that maximise radiation intensity while keeping the peak photon energy stable.

Importantly, these results are consistent with the scaling behaviour predicted by the continuous potential model in \Cref{eqn:First_Harmonic}, demonstrating that it provides a reliable estimate for the positions of CUR peaks over a broad range of bending properties. The atomistic MD simulations extend these predictions by explicitly modelling discrete particle-atom interactions such as elastic and inelastic scattering, which govern dechannelling and rechannelling processes. While these effects have a minor influence on the CUR peak positions, they directly affect channelling dynamics, efficiency, and radiation intensity. This is directly observed from the decrease in spectral intensity beyond $a=\SI{1.44}{\angstrom}$ for $\lambda=\SI{3.5}{\micro\meter}$, where the bending parameter $C$ is large, and the increased dechannelling suppresses the radiation intensity, showing the limits of the continuous potential approximation.

Simulations of a four-period crystal undulator with parameters matching those used at MAMI predict a CUR peak at approximately \SI{0.515}{\mega\electronvolt} for \SI{530}{\mega\electronvolt} positrons. Overall, the position of the CUR peak remains stable across a broad range of bending amplitudes and periods: small variations in the bending amplitude have little effect, while the peak is more sensitive to changes in the bending period.

These results provide a framework for estimating CUR peak positions in future channelling experiments. Although this study has focused on \SI{530}{\mega\electronvolt} positrons at MAMI, the same methodology and trends are applicable to higher-energy electron channelling, the planned \SI{855}{\mega\electronvolt} positron beamline extension at MAMI, and other accelerator facilities such as CERN.

\section*{Statements and declarations}

\bmhead{Acknowledgments}
\noindent This work was funded by UK Research and Innovation (UKRI) under the UK government’s Horizon Europe Funding Guarantee (grant No.~10037865), in collaboration with the European Commission's Horizon Europe EIC-Pathfinder-Open TECHNO-CLS project (G.A.~101046458).
The authors also acknowledge financial support from the H2020 RISE-NLIGHT
project (G.A.~872196).
The work was supported in part by Deutsche Forschungsgemeinschaft, Germany (Project No.~413220201).
The authors gratefully acknowledge the Specialist and High Performance Computing systems provided by Information Services at the University of Kent. Special thanks are extended to Dr Timothy Kinnear for HPC assistance with the Icarus cluster at the University of Kent.
The authors kindly thank the anonymous referee, whose valuable comments have improved the quality and clarity of this manuscript.

\bmhead{Competing Interests}
The authors do not declare any conflicts of interest, and there is no financial interest to report.

\bmhead{Author Contribution}
MDD and AVK conducted the computational simulations and data analysis. MDD wrote the manuscript with support from AVK and input from all authors. NJM, FF, and AVS helped supervise the project and all authors provided critical feedback and helped shape the research and analysis.

\bmhead{Data Availability}
The datasets generated and/or analysed during the current study are available from the corresponding author upon reasonable request.

\bibliography{sn-bibliography}

\clearpage
\setcounter{section}{0}
\setcounter{page}{1}
\setcounter{figure}{0}
\setcounter{equation}{0}

\renewcommand{\thesection}{S\arabic{section}}
\renewcommand{\thepage}{S\arabic{page}}
\renewcommand{\thetable}{S\arabic{table}}
\renewcommand{\thefigure}{S\arabic{figure}}
\renewcommand{\theequation}{S\arabic{equation}}

\onecolumn

\begin{center}
\large Supplemental Information for:\\Influence of bending parameters on crystalline undulator radiation peak stability for 530~MeV positron channelling
\end{center}

\section{Tables of Physical Parameters} \label{app:Parameters_Table}
The following tables contain the values associated with the key physical parameters for each amplitude-period combination considered in this study.

\begin{table}[ht]
\centering
\caption{Bending Parameter $C$}
\label{tab:Bending_Parameter}
\begin{tabular}{c|cccccccccc}
\hline
\multirow{2}{*}[0pt]{\shortstack{Bending\\Amplitude $a$ (\unit{\angstrom})}} &
\multicolumn{10}{c}{Bending Period $\lambda$ (\unit{\micro\meter})} \\
& 3.5 & 4.0 & 4.5 & 5.0 & 5.5 & 6.0 & 6.5 & 7.0 & 7.5 & 8.0 \\ \hline
\multicolumn{1}{c|}{0.60}  & 0.171 & 0.131 & 0.103 & 0.084 & 0.069 & 0.058 & 0.050 & 0.043 & 0.037 & 0.033 \\
\multicolumn{1}{c|}{0.87} & 0.231 & 0.177 & 0.139 & 0.113 & 0.093 & 0.078 & 0.067 & 0.058 & 0.050 & 0.044 \\
\multicolumn{1}{c|}{1.02} & 0.290 & 0.222 & 0.176 & 0.142 & 0.118 & 0.099 & 0.084 & 0.073 & 0.063 & 0.056 \\
\multicolumn{1}{c|}{1.23} & 0.350 & 0.268 & 0.212 & 0.172 & 0.142 & 0.199 & 0.102 & 0.088 & 0.076 & 0.067 \\
\multicolumn{1}{c|}{1.44} & 0.410 & 0.314 & 0.248 & 0.201 & 0.166 & 0.139 & 0.119 & 0.102 & 0.089 & 0.078 \\
\multicolumn{1}{c|}{1.66} & 0.473 & 0.362 & 0.286 & 0.232 & 0.191 & 0.161 & 0.137 & 0.118 & 0.103 & 0.090 \\
\multicolumn{1}{c|}{1.87} & 0.532 & 0.408 & 0.322 & 0.261 & 0.216 & 0.181 & 0.154 & 0.133 & 0.116 & 0.102 \\
\multicolumn{1}{c|}{2.08} & 0.592 & 0.453 & 0.358 & 0.290 & 0.240 & 0.201 & 0.172 & 0.148 & 0.129 & 0.113 \\
\multicolumn{1}{c|}{2.29} & 0.652 & 0.499 & 0.394 & 0.319 & 0.264 & 0.222 & 0.189 & 0.163 & 0.142 & 0.125 \\
\multicolumn{1}{c|}{2.50} & 0.712 & 0.545 & 0.431 & 0.349 & 0.288 & 0.242 & 0.206 & 0.178 & 0.155 & 0.136
\end{tabular}
\end{table}

\begin{table}[ht]
\centering
\caption{Undulator Parameter $K_{\text{u}}^2$}
\label{tab:Undulator_Parameter_Ku}
\begin{tabular}{c|cccccccccc}
\hline
\multirow{2}{*}[0pt]{\shortstack{Bending\\Amplitude $a$ (\unit{\angstrom})}} &
\multicolumn{10}{c}{Bending Period $\lambda$ (\unit{\micro\meter})} \\
& 3.5 & 4.0 & 4.5 & 5.0 & 5.5 & 6.0 & 6.5 & 7.0 & 7.5 & 8.0 \\ \hline
\multicolumn{1}{c|}{0.60}  & 0.112 & 0.098 & 0.087 & 0.078 & 0.071 & 0.065 & 0.060 & 0.056 & 0.052 & 0.049 \\
\multicolumn{1}{c|}{0.87} & 0.151 & 0.132 & 0.117 & 0.106 & 0.096 & 0.088 & 0.081 & 0.075 & 0.070 & 0.066 \\
\multicolumn{1}{c|}{1.02} & 0.190 & 0.166 & 0.148 & 0.133 & 0.121 & 0.111 & 0.102 & 0.095 & 0.089 & 0.083 \\
\multicolumn{1}{c|}{1.23} & 0.229 & 0.200 & 0.178 & 0.160 & 0.146 & 0.134 & 0.123 & 0.115 & 0.107 & 0.100 \\
\multicolumn{1}{c|}{1.44} & 0.268 & 0.235 & 0.209 & 0.188 & 0.171 & 0.156 & 0.144 & 0.134 & 0.125 & 0.117 \\
\multicolumn{1}{c|}{1.66} & 0.309 & 0.270 & 0.240 & 0.216 & 0.197 & 0.180 & 0.166 & 0.155 & 0.144 & 0.135 \\
\multicolumn{1}{c|}{1.87} & 0.348 & 0.305 & 0.271 & 0.244 & 0.222 & 0.203 & 0.187 & 0.174 & 0.162 & 0.152 \\
\multicolumn{1}{c|}{2.08} & 0.387 & 0.339 & 0.301 & 0.271 & 0.246 & 0.226 & 0.209 & 0.194 & 0.181 & 0.169 \\
\multicolumn{1}{c|}{2.29} & 0.426 & 0.373 & 0.332 & 0.298 & 0.271 & 0.249 & 0.230 & 0.213 & 0.199 & 0.187 \\
\multicolumn{1}{c|}{2.50} & 0.465 & 0.407 & 0.362 & 0.326 & 0.296 & 0.272 & 0.251 & 0.233 & 0.217 & 0.204
\end{tabular}
\end{table}

\begin{table}[ht]
\centering
\caption{Combined Undulator Parameter $K^2=K_{\text{u}}^2+K_{\text{ch}}^2(1-C)$}
\label{tab:Combined_Undulator_K}
\begin{tabular}{c|cccccccccc}
\hline
\multirow{2}{*}[0pt]{\shortstack{Bending\\Amplitude $a$ (\unit{\angstrom})}} &
\multicolumn{10}{c}{Bending Period $\lambda$ (\unit{\micro\meter})} \\
& 3.5 & 4.0 & 4.5 & 5.0 & 5.5 & 6.0 & 6.5 & 7.0 & 7.5 & 8.0 \\ \hline
\multicolumn{1}{c|}{0.60}  & 0.035 & 0.033 & 0.032 & 0.031 & 0.030 & 0.030 & 0.029 & 0.029 & 0.029 & 0.029 \\
\multicolumn{1}{c|}{0.87} & 0.044 & 0.040 & 0.037 & 0.035 & 0.034 & 0.033 & 0.032 & 0.031 & 0.031 & 0.030 \\
\multicolumn{1}{c|}{1.02} & 0.055 & 0.049 & 0.044 & 0.041 & 0.038 & 0.037 & 0.035 & 0.034 & 0.033 & 0.032 \\
\multicolumn{1}{c|}{1.23} & 0.070 & 0.060 & 0.053 & 0.048 & 0.044 & 0.042 & 0.040 & 0.038 & 0.036 & 0.035 \\
\multicolumn{1}{c|}{1.44} & 0.088 & 0.074 & 0.064 & 0.057 & 0.052 & 0.048 & 0.045 & 0.042 & 0.040 & 0.039 \\
\multicolumn{1}{c|}{1.66} & 0.110 & 0.090 & 0.077 & 0.068 & 0.061 & 0.055 & 0.051 & 0.048 & 0.045 & 0.043 \\
\multicolumn{1}{c|}{1.87} & 0.134 & 0.109 & 0.092 & 0.079 & 0.070 & 0.063 & 0.058 & 0.054 & 0.050 & 0.048 \\
\multicolumn{1}{c|}{2.08} & 0.161 & 0.130 & 0.108 & 0.093 & 0.081 & 0.073 & 0.066 & 0.061 & 0.056 & 0.053 \\
\multicolumn{1}{c|}{2.29} & 0.191 & 0.153 & 0.126 & 0.108 & 0.094 & 0.083 & 0.075 & 0.068 & 0.063 & 0.058 \\
\multicolumn{1}{c|}{2.50} & 0.224 & 0.178 & 0.146 & 0.124 & 0.107 & 0.094 & 0.084 & 0.076 & 0.070 & 0.065
\end{tabular}
\end{table}

\begin{table}[ht]
\centering
\caption{First Harmonic Position $\hbar\omega_1$}
\label{tab:First_Harmonic_Position}
\begin{tabular}{c|cccccccccc}
\hline
\multirow{2}{*}[0pt]{\shortstack{Bending\\Amplitude $a$ (\unit{\angstrom})}} &
\multicolumn{10}{c}{Bending Period $\lambda$ (\unit{\micro\meter})} \\
& 3.5 & 4.0 & 4.5 & 5.0 & 5.5 & 6.0 & 6.5 & 7.0 & 7.5 & 8.0 \\ \hline
\multicolumn{1}{c|}{0.60}  & 0.749 & 0.656 & 0.584 & 0.526 & 0.478 & 0.438 & 0.405 & 0.376 & 0.351 & 0.329 \\
\multicolumn{1}{c|}{0.87} & 0.746 & 0.654 & 0.582 & 0.524 & 0.477 & 0.438 & 0.404 & 0.375 & 0.350 & 0.329 \\
\multicolumn{1}{c|}{1.02} & 0.742 & 0.651 & 0.580 & 0.523 & 0.476 & 0.437 & 0.403 & 0.375 & 0.350 & 0.328 \\
\multicolumn{1}{c|}{1.23} & 0.737 & 0.648 & 0.578 & 0.521 & 0.475 & 0.436 & 0.403 & 0.374 & 0.349 & 0.328 \\
\multicolumn{1}{c|}{1.44} & 0.730 & 0.643 & 0.575 & 0.519 & 0.473 & 0.434 & 0.402 & 0.373 & 0.349 & 0.327 \\
\multicolumn{1}{c|}{1.66} & 0.723 & 0.638 & 0.571 & 0.516 & 0.471 & 0.433 & 0.400 & 0.372 & 0.348 & 0.327 \\
\multicolumn{1}{c|}{1.87} & 0.715 & 0.633 & 0.567 & 0.513 & 0.469 & 0.421 & 0.399 & 0.371 & 0.347 & 0.326 \\
\multicolumn{1}{c|}{2.08} & 0.706 & 0.627 & 0.563 & 0.510 & 0.466 & 0.429 & 0.397 & 0.370 & 0.346 & 0.325 \\
\multicolumn{1}{c|}{2.29} & 0.696 & 0.620 & 0.558 & 0.506 & 0.464 & 0.427 & 0.396 & 0.369 & 0.345 & 0.324 \\
\multicolumn{1}{c|}{2.50} & 0.686 & 0.631 & 0.553 & 0.503 & 0.461 & 0.425 & 0.394 & 0.367 & 0.344 & 0.323
\end{tabular}
\end{table}

\begin{table}[ht]
\centering
\caption{Acceptance $\mathcal{A}$}
\label{tab:Acceptance}
\begin{tabular}{c|cccccccccc}
\hline
\multirow{2}{*}[0pt]{\shortstack{Bending\\Amplitude $a$ (\unit{\angstrom})}} &
\multicolumn{10}{c}{Bending Period $\lambda$ (\unit{\micro\meter})} \\
& 3.5 & 4.0 & 4.5 & 5.0 & 5.5 & 6.0 & 6.5 & 7.0 & 7.5 & 8.0 \\ \hline
\multicolumn{1}{c|}{0.60} & 0.858 & 0.884 & 0.904 & 0.920 & 0.924 & 0.934 & 0.935 & 0.938 & 0.933 & 0.938 \\
\multicolumn{1}{c|}{0.87} & 0.808 & 0.855 & 0.883 & 0.910 & 0.921 & 0.930 & 0.915 & 0.928 & 0.933 & 0.936 \\
\multicolumn{1}{c|}{1.02} & 0.790 & 0.833 & 0.880 & 0.889 & 0.884 & 0.904 & 0.920 & 0.914 & 0.920 & 0.927 \\
\multicolumn{1}{c|}{1.23} & 0.727 & 0.789 & 0.849 & 0.870 & 0.888 & 0.899 & 0.905 & 0.910 & 0.912 & 0.921 \\
\multicolumn{1}{c|}{1.44} & 0.678 & 0.763 & 0.807 & 0.845 & 0.868 & 0.880 & 0.893 & 0.910 & 0.916 & 0.915 \\
\multicolumn{1}{c|}{1.66} & 0.621 & 0.739 & 0.787 & 0.819 & 0.854 & 0.872 & 0.884 & 0.894 & 0.907 & 0.904 \\
\multicolumn{1}{c|}{1.87} & 0.581 & 0.692 & 0.763 & 0.791 & 0.826 & 0.850 & 0.875 & 0.893 & 0.882 & 0.892 \\
\multicolumn{1}{c|}{2.08} & 0.536 & 0.659 & 0.748 & 0.776 & 0.810 & 0.848 & 0.851 & 0.880 & 0.888 & 0.890 \\
\multicolumn{1}{c|}{2.29} & 0.509 & 0.605 & 0.699 & 0.754 & 0.804 & 0.827 & 0.850 & 0.867 & 0.873 & 0.883 \\
\multicolumn{1}{c|}{2.50} & 0.440 & 0.599 & 0.697 & 0.737 & 0.779 & 0.800 & 0.859 & 0.852 & 0.865 & 0.879
\end{tabular}
\end{table}

\end{document}